\documentclass[oldversion]{aa}
\usepackage{graphicx}
\usepackage{natbib}
\bibpunct{(}{)}{;}{a}{}{,}
\usepackage{txfonts}

\begin{document}

\newcommand{\micron}{\rm \mu m}
\newcommand{\Snu}{S_\nu}
\newcommand{\Bnu}{B_\nu}
\newcommand{\eexp}{\rm e}
\newcommand{\tauff}{\tau_{\rm ff}}
\newcommand{\gaunt}{g_{\rm ff}}
\newcommand{\nel}{n_{\rm e}}

\title{High-Spatial Resolution SED of NGC~1068 from Near-IR to Radio}
\subtitle{Disentangling the thermal and non-thermal contributions}

\author{S.~F.~H\"onig\inst{1} \and
M.~A.~Prieto\inst{2} \and
T.~Beckert\inst{1}}

\offprints{S.~F. H\"onig \\ \email{shoenig@mpifr-bonn.mpg.de}}

\institute{Max-Planck-Institut f\"ur Radioastronomie, Auf dem H\"ugel 69, 53121 Bonn, Germany \and
			Instituto Astrofisica de Canarias, 38200 La Laguna, Tenerife, Spain}

\date{Received ... / Accepted ...}

  \abstract{
We investigate the ideas that a sizable fraction of the interferometrically unresolved infrared emission of the nucleus of NGC 1068 might originate from other processes than thermal dust emission from the torus. We examine the contribution of free-free or synchrotron emissions to the central mid- and near-IR parsec-scale emitting region of NGC 1068. Each mechanism is constrained with parsec scale radio data available for NGC 1068 in the $10^9 -10^{11}$\,Hz regime, and compared to the highest-resolution interferometric data available in the mid-infrared. It is shown that the unresolved emission in the interferometric observation ($\la$1\,pc) is still dominated by dust emission and not by contributions from synchrotron or free-free emission. As recent studies suggest, the interferometric observations prefer a clumpy structure of the dust distribution. Extrapolation of the radio free-free or synchrotron emission to the IR indicates that their contribution is $<$20\% even for the unresolved fraction of the interferometric flux.

The slope of the available radio data is consistent with a power law exponent $\alpha = 0.29 \pm 0.07$ which we interprete in terms of either free-free emission or synchrotron radiation from quasi-monochromatic electrons. We apply emission models for both mechanisms in order to obtain physical parameters.

Furthermore, we aimed for quantifying the possible contribution of the accretion disk to the near-infrared emission. It has been suggested, that the unresolved $K$-band flux in VLTI/VINCI interferometric observation at 46\,m baseline (40\% of the total $K$-band flux) might originate from the accretion disk. By using an accretion disk spectrum which was adjusted to the luminosity and black hole mass of NGC 1068, we found out that the expected accretion disk flux in the $K$-band is negligible. Moreover, the scenario of detecting the accretion disk through holes in a clumpy torus is extremely unlikely. We conclude that all current IR data of NGC 1068 trace the torus dust emission, favouring a clumpy torus.


  }

   \maketitle
%

\section{Introduction} \label{Intro}

The nucleus of the nearby Seyfert 2 galaxy NGC~1068 ($d\sim14.4\,{\rm Mpc}$; a scale of 10\,mas corresponds to $\sim$0.7\,pc) is the prime target of the most recent high-spatial resolution studies of AGN in the wavelength regime from infrared (IR) to the radio \citep[e.g.,][]{Wei04,Wit04,Jaf04,Gal04}. For the first time, it is possible to build-up a genuine nuclear spectrum of the immediate vicinity of the AGN. Furthermore, near- (NIR) and mid-infrared (MIR) interferometric observations resolved the nulcear IR emission source \citep{Wit98,Wei04,Jaf04}. It is supposed that the nuclear radio emission comes from an accretion disk corona or an associated jet while the infrared emission is more likely caused by thermal dust emission in the circumnuclear dusty torus \citep[e.g.,][]{Gal96,Gal04,Hon06}.

It is, however, yet unclear how much each component contributes at different wavelengths. MIR VLTI/MIDI interferometric observations by \citet{Jaf04} show that approximately 20\% of the MIR emission originates from an unresolved source $\la0.7\,{\rm pc}$. \citet{Wit04} report that $K$-band interferometry with the VLTI/VINCI instrument at 46\,m baseline reveals a surprisingly high contribution of 40\% to the whole nuclear flux by an unresolved source of $\la0.3\,{\rm pc}$. This can be explained by either radiation from a small compact source which is associated to the nuclear radio spectrum \citep[e.g.,][]{Wit98} or by substructure inside a clumpy dust torus \citep{Hon06}.

In this paper, we use the combination of high-spatial resolution photometry, radiative-transfer simulations in the IR, and emission models in the radio to compare the extrapolated radio emission to measurements at IR wavelengths, in order to investigate the origin of the resolved and unresolved the near- and mif-infrared emission. In Sect.~\ref{Obs}, we summarize the observational data used for our study. In Sect.~\ref{Models}, we describe the models which were used for the radio and IR regimes, respectively. In the following Sect.~\ref{Comp}, we discuss the model results and analyse the maximum contributions of the radio component to the observed NIR and MIR emission of diffraction-limited and interferometric data.
In Sect.~\ref{AD}, the contribution of the accretion disk to the IR spectrum is quantified and compared to the interferometric VLTI/VINCI observations. In Sect.~\ref{Conc}, we summarize our results.

\section{Observational data}\label{Obs}

\begin{table}
\caption{High spatial resolution observations of the nulear emission of NGC\,1068 from the radio to the infrared regime.}\label{tab1}
\centering
\begin{tabular}{c c c l}
\hline\hline
Frequency & 		Flux & 				spatial scale &			Reference \\
 & [Jy] & [mas] & \\ \hline
$2.3\times10^{14}$ &	0.0084 & 			78 &					this work \\
$1.8\times10^{14}$ &	0.022 &				130 &					this work \\
$1.4\times10^{14}$ &	0.098 &				78 &					this work \\
$1.2\times10^{14}$ & $0.32\pm0.03$&		78 &					this work \\
$7.9\times10^{13}$ & 1.5 & 				700 &					[1] \\
$6.7\times10^{13}$ & 2.5 &					162 &					this work \\
$3.8\times10^{13}$ & 4.23 &				200 &					[2] \\
$3.0\times10^{13}$ & 	3.33 &				200 &					[2] \\
$2.5\times10^{13}$ & 	7.1 &				200 &					[2] \\
$1.6\times10^{13}$ & 	9.6 &				$290\times180$ &			[3] \\
$1.2\times10^{13}$ & 	9.4 &				200 &					[2] \\
$2.3\times10^{11}$ & 	$0.022\pm0.008$ &		$1000\times600$ &			[4] \\
$1.15\times10^{11}$ &	$0.023\pm 0.05$ &		1400 &					[4] \\
$4.3\times10^{10}$ &	$0.019 \pm 0.003$ &	75 &					[5] \\
$2.25\times10^{10}$ & $0.016 \pm 0.006$ &	75 &					[5] \\
$8.4\times10^{9}$ &	$0.0054 \pm 0.0005$ &	$11.0\times3.7$ &			[6] \\
$5.0\times10^{9}$ &	$0.0091 \pm 0.0008$ &	$16.6\times11.2$ &			[6] \\
$5.0\times10^{9}$ &	0.0051 &			$14\times3$ &				[7] \\
$1.7\times10^{9}$ &	$<0.0026$ &			$14\times12$ &			[7] \\
$1.4\times10^{9}$ &	$<6\times10^{-5}$ &	$16.0\times7.6$ &			[6] \\ \hline
\multicolumn{4}{l}{[1] \citet{Mar03}; [2] \citet{Boc00};}\\
\multicolumn{4}{l}{[3] \citet{Tom01}; [4] \citet{Kri06};}\\
\multicolumn{4}{l}{[5] \citet{Mux96}; [6] \citet{Gal04};}\\
\multicolumn{4}{l}{[7] \citet{Roy98}}
\end{tabular}
\end{table}

The location of NGC 1068's nucleus in the radio is assumed to be at the center of a 0.8\,pc size, disk-like structure known as S1, resolved at 5 and 8.4\,GHz by \citet{Gal97,Gal04}. In the mid-IR, a gaussian decomposition of VLTI/MIDI interferometry data sets an upper limit of 0.7\,pc $\times$ $<$1\,pc on the size of the central source at wavelengths $<10\,\micron$ \citep{Jaf04}. At $2\,\micron$, the combination of speckle observations \citep{Wei04} and VLTI/VINCI interferometry \citep{Wit04} confirm the presence of a compact core with FWHM $\sim$0.4\,pc. Shortward of $1\,\micron$, the NGC 1068 nucleus is fully obscured and undetected.

The radio to IR to optical SED of N1068's nucleus is extracted from subarcsec photometry currently available for this source. Overall, we sample spatial scales at radio, MIR and NIR wavelengths down to few parsecs. Radio data for the source S1, which is assumed to be NGC 1068 nucleus, is taken from the VLA/MERLIN 65 mas beam \citep{Mux96,Gal04} and VLBA-observation \citep{Roy98,Gal04}. \citet{Gal04} report that they resolved the nuclear source S1 at 5 and 8.4\,GHz. S1 appears disk-like with a reference size of $\sim$11\,mas (for detailed sizes see Table~\ref{tab1}). The millimeter data at 1\,mm and 3\,mm are taken from \citet{Kri06} and correspond to a beam size of $1\times0.6$\,arcsec. Because of the possible contribution of the jet in the 3\,mm data, only the peak value as specified by \citet{Kri06} is considered.

Data in the $8-25\,\micron$ range are taken from \citet{Boc00} and \citet{Tom01}, using their reported fluxes in aperture diameters of $\sim$200\,mas. This high resolution was obtained by deconvolving their 10\,m Keck and 8.2\,m Subaru images, respectively. It is important to note that both sets of deconvolved-extracted fluxes differ by $<$30\%.

Interferometry data from VLTI/MIDI in the $8-13\,\micron$ window are taken from \citet{Jaf04}. For our study, we used the correlated fluxes from the 78\,m baseline which provides the highest resolution data currently available. The visibilities show that the silicate absorption feature becomes more pronounced with growing baseline. This was interpreted as an indication for clumpy substructure within the dust torus \citep{Jaf04,Hon06}. 5-20\% of the total flux are unresolved at the 78\,m baseline.

In the $1-5\,\micron$ range\footnote{The near-IR wavebands are referred to as $J$-band ($1.3\,\micron \Leftrightarrow 2.3\times10^{14}$\,Hz), $H$-band ($1.65\,\micron \Leftrightarrow 1.8\times10^{14}$\,Hz), $K$-band ($2.2\,\micron \Leftrightarrow 1.4\times10^{14}$\,Hz), $L$-band ($3.8\,\micron \Leftrightarrow 7.9\times10^{13}$\,Hz), and $M$-band ($4.5\,\micron \Leftrightarrow 6.7\times10^{13}$\,Hz). Mid-IR wavebands are referred to as $N$-band ($8-13\,\micron \Leftrightarrow\, \sim\!3\times10^{13}$\,Hz) and $Q$-band ($16-25\,\micron \Leftrightarrow\, \sim\!1.5\times10^{13}$\,Hz).}, we use our adaptive optics images acquired with VLT/NACO, except for the $L$-band photometry which was taken from \citet{Mar03}. The spatial resolution achieved in these observations varies from FWHM$\sim$78\,mas in the $1.3-2.44\,\micron$ range to FWHM$\sim$160\,mas in the $M$-band. Nuclear fluxes were integrated in an aperture size comparable to those resolutions. These aperture sizes were selected as a compromise to minimize the contribution from the prominent IR diffuse emission surrounding the nucleus of NGC 1068. The $1.3\,\micron$ measurement should be considered as an upper limit. The corresponding frequency is the highest at which the core of NGC 1068 can be reliably distinguished from the strong extended emission.

NGC 1068 is Compton thick in the X-ray regime. It shows prominent extended soft X-ray emission as seen by {\it Chandra} \citep{You01}. The extended emission is preferentially distributed along the ionization cone. XMM spectra show that most of this emission is resolved in multiple emission lines and not much continuum is detected \citep{Kin02}. Due to the Compton thickness of the source and extension of the emission, we do not include X-ray photometry in our modelling of the nucleus.

\section{Models}\label{Models}

We focused our modelling efforts on two components: (1) the near- and mid-infrared component in the range of $10^{13}-10^{14}$\,Hz, and (2) the radio component at frequencies from $10^{9}$\,Hz to $10^{11}$\,Hz. Our aim is to quantify the contribution of each component in different wavelength regimes.

\subsection{IR torus model}\label{IRModel}

\begin{figure*}
\centering
\includegraphics[angle=0,width=14.5cm]{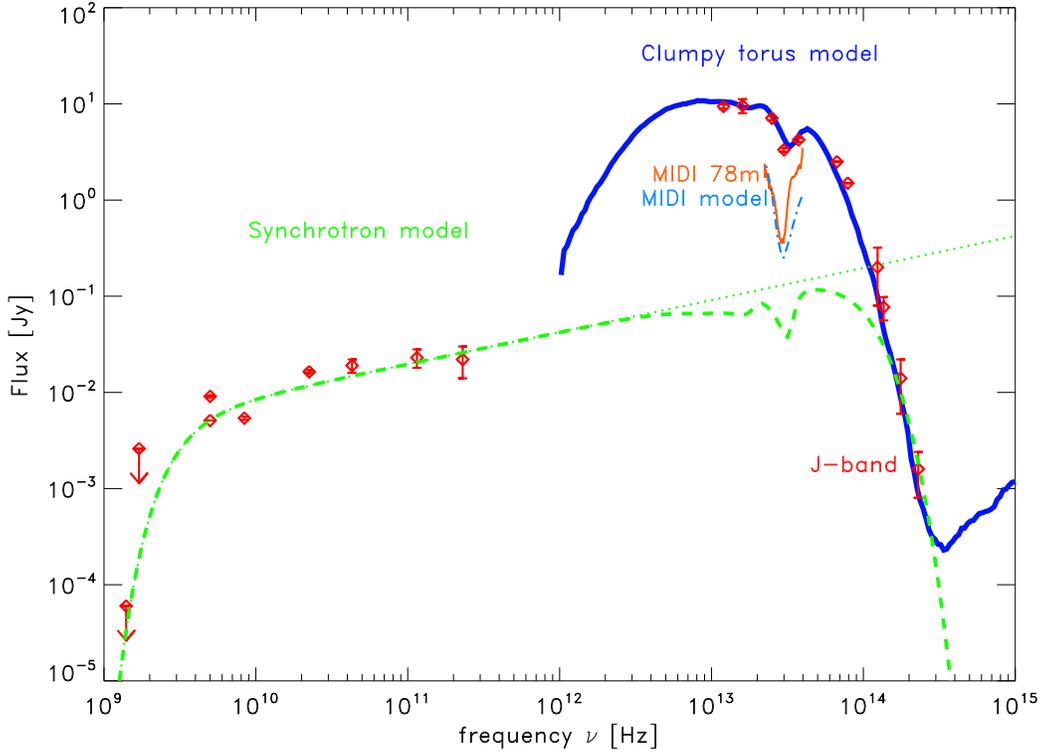}
\caption{High-spatial resolution SED of the nucleus of NGC~1068, in comparison to thermal dust emission from a torus in the IR and non-thermal synchrotron emission in the radio. The blue line shows the clumpy torus model that was used in this study to model the single-dish IR photometry (diamonds) and interferometric observations by VLTI/MIDI (orange solid line; model: dashed-dotted blue line). The radio data (diamonds) are modelled by a spectrum $F_\nu \propto \nu^{1/3}\exp\left(-(\nu/\nu_{\rm ff})^{-2.1}\right)$ (Synchrotron emission + free-free absorption; green-dotted line) and extrapolated into the infrared. The dashed line shows the synchrotron spectrum attenuated by dust obscuration ($\tau_V=16.8$) to reproduce the $J$-band data.
regimes.}\label{Sync}
\end{figure*}

For the infrared regime of the NGC~1068 SED, we used the clumpy torus models of \citet{Hon06}. With these models, we aimed at fitting the photometric data from $1.2-23\times10^{13}$\,Hz, simultaneously with the 78\,m baseline data as observed with VLTI/MIDI. A grid of models has been calculated where we tested (a) different radial cloud distributions (radial distribution function $\propto r^{-a}$ with $a=1.0,1.5,2.0$), (b) inclincation angles, and (c) average numbers of (optically thick) clouds, $N_0$, in an equatorial line of sight. For each set of model parameters we simulated 10 different random arrangements of clouds. This accounts for the variations in SEDs which occur due to the actual cloud distribution. Our best-fitting model is shown as blue-dashed lines in Figs.~\ref{Sync} \& \ref{Free}.

\begin{table*}
\caption{Parameters of the synchrotron models A, B, and C for the nuclear source S1 of NGC~1068: The linear scale $s$ is the half thickness in model A \& B and the radius of the emission spheres in model C. For the mean Lorentz factor $\left<\gamma\right>$ of the electron distributions (width $\Delta \gamma /\left<\gamma\right> = 2$) there are only lower limits derived form the assumed upper cut-off frequency $\nu_c = 3\times10^{14}$\,Hz of the synchrontron spectrum. The values for $B_\mathrm{Min}$ is the minimum energy $B$-field for the claimed size of $s=5.6$\,mas in model A and upper limits for B and C. The electron density in the last column corresponds to $B_\mathrm{Min}$.}              
\label{tableSync}      
\centering                           
\begin{tabular}{clllll}        
\hline\hline                 
Model & $s$ &  $s \times d$ & $\left<\gamma\right>$ & $B_\mathrm{Min} $ & $\nel$ \\     
& [mas] & [pc] &  & [G] & [cm$^{-3}$] \\
\hline                        
 A & $5.6$ & $0.39$ & $>1.5\times10^5$ & $1.6\times10^{-2}$  & $1.3\times10^{-4} $\\       
 B & $>2\times10^{-7}$ & $>1.3\times10^{-8}$ & $>1.3\times10^4$ & $<2.2$ & $<3.0 $\\       
 C &  $>1.6\times10^{-3}$ & $>1.1\times10^{-4}$ & $>8\times10^3$ & $<6.1$ & $<350 $ \\
\hline                                    
\end{tabular}
\end{table*}

The modeled torus has a line-of-sight inclination of 90\,deg, which corresponds to an edge-on view of the torus. In radial direction, the clouds are distributed according to a power-law of $r^{-3/2}$, which is well consistent with clumpy torus accretion scenarios where the torus is located in the sphere of influence of the central supermassive black hole \citep{Bec04}. We assumed linear flaring throughout the torus, i.e. the height of the torus $H$ increases with radius, $H \propto r$. The half opening angle of our geometrical configuration is $\approx$40\,deg. Although a general power-law $H \propto r^x$ could have been chosen, it has already been tested in \citet{Hon06,Hon07a} and no better fit to the IR data was found then. Furthermore, theoretical considerations and observational studies of clumpy tori seem to favour linear flaring geometries over stronger $r$-dependences of the torus height \citep{Hon07b,Pol08}. The obscuration of the nucleus in the presented torus model is characterised by an average of $N_0=10$ optically thick dust clouds in an equatorial line of sight. For the edge-on geometry that we modelled, this corresponds to average number of clouds, $N$, which obscure the AGN along the line of sight towards the observer (i.e., $N=N_0=10$, as used in Sect.~\ref{AD}).

As compared to other model parameters in our grid, it simultaneously reproduces (1) the NIR slope, (2) the depth of the silicate feature, (3) the slope from $N$-band to $Q$-band, and (4) the correlated MIDI flux. Due to the ambiguity in actual cloud arrangement, there is some degeneracy in the model parameter space \citep[compare results in][]{Hon06}, especially when fitting the SED alone. On the other hand, the presented dust distribution parameters are the only one for which we found good agreement with the SED properties (1 -- 3) {\it and} the interferometric data (4). As a general guideline, $a>1.5$ produces bluer NIR colors and less $Q$-band flux, and vice versa for $a<1.5$ \citep[see also discussion in][]{Pol08}.

The observed and modeled diameter of the dust torus is $\sim$20-30\,mas ($\sim$1.4-2\,pc), depending on wavelength and orientation. To fit the observed fluxes, a bolometric luminosity of the central AGN of $2\times10^{45}\,{\rm erg/s}$ has been used. We note that this model luminosity is larger than inferred from the measured fluxes assuming isotropic emission. The reasons for this are actual anisotropic emission of the torus (type 2 edge-on emission is smaller than type 1 face-on emission) as well as the covering factor of the torus which is $<1$, so that not all of the AGN bolometric luminosity is absorbed by the torus. Total discrepancies can be up to one order of magnitude.

\subsection{Radio models}

In the radio regime at frequencies lower than 1 THz, we used two different emission models to reproduce the observed high-spatial resolution data. In order to study the possible influence of the radio emission sources to the observed IR emission, we will assume that the model spectra reach the near-IR regime without a high-frequency break. In a further step, the model spectra are attenuated by dust extinction so that the model fluxes are consistent with the observed near-IR fluxes. For that, we use different optical depths, $\tau_V$, so that the observed $J$-band flux at 1.2\,$\micron$ is reproduced only by the radio component (i.e., 100\% flux contribution). This procedure will give us a very strong upper limit to the possible contribution of the radio source to the $K$-band data where high-spatial resolution observations and interferometry is available. We will compare the used $\tau_V$ values with the torus model and the estimated broad-line region obscuration in the next section.

\paragraph{Synchrotron radiation.} The radio data of NGC\,1068 shows an inverted spectrum ($\alpha > 0$) between 5-230\,GHz with an apparent turnover at $\sim10^9$\,Hz. From a least-square fit to actual radio detections we obtain a slope of $\alpha=0.29\pm0.07$. This is consistent with a power law index $\alpha_{\rm mono}=1/3$, which can be produced by optically thin synchrotron radiation coming from quasi-monoenergetic electrons \citep{Bec97}. In such a scenario, the emission spectrum can be approximated by
\begin{equation}
\Snu \propto \nu^{1/3}\exp(-\nu/\nu_c) \
\end{equation}
with an exponential turnover, $\nu_c$, at high frequencies. For the purpose of estimating the maximum possible NIR contribution of the radio source, we assume that $\nu_c\ga300$\,THz, so that the actual high-frequency cutoff can be neglected hereafter. At low frequencies, a turnover can occur, e.g. if the synchrotron source is located within dense plasma leading to thermal free-free absorption $\tau_{\rm ff} \propto \nu^{-2.1}$ which produces a sharp turnover towards low frequencies \citep[e.g.,][]{Bec96}. As a result, the synchrotron spectrum is modified by a factor $\exp\left(-(\nu/\nu_{\rm ff})^{-2.1}\right)$, with the low-frequency turnover at $\nu_{\rm ff}$, so that the flux can be parameterised by
\begin{equation}
\Snu \propto \nu^{1/3} \exp\left(-(\nu/\nu_{\rm ff})^{-2.1}\right) \ ,
\end{equation}
neglecting the $\exp(-\nu/\nu_c)$ turnover at high frequencies.

In Fig.~\ref{Sync}, we compare the synchrotron model SED with observed radio fluxes. The low-frequency free-free absorption turnover occurs at $\nu_{\rm ff}=3$\,GHz which is consistent with the upper limit on detections around 1.5\,GHz. The spectrum was attentuated by standard ISM dust absorption of optical depth $\tau_V=16.8$ to reproduce the $J$-band observation.

An alternative explanation for the low-frequency turnover is synchrotron self-absorption (SSA) within the source. This would result in a spectral turnover from the optically thin $F_\nu \propto \nu^{1/3}$ to the optically thick case $F_\nu \propto \nu^{2}$, which is less steep than in case of free-free absorption as discussed above. The optically thick part behaves like a thermal Rayleigh-Jeans branch in contrast to the standard power-law electron distribution which produces $F_\nu \propto \nu^{5/2}$ \citep[e.g.,][]{Pach70}. Given the detection of S1 in NGC~1068 at 5\,GHz and the upper limits at $\sim$1.4\,GHz, the turnover should occur inbetween. The observed lower limit for the power law index in this frequency range ($\alpha \ga$3.5) favours the free-free absorption scenario over SSA.

\begin{figure*}
\centering
\includegraphics[angle=0,width=14.5cm]{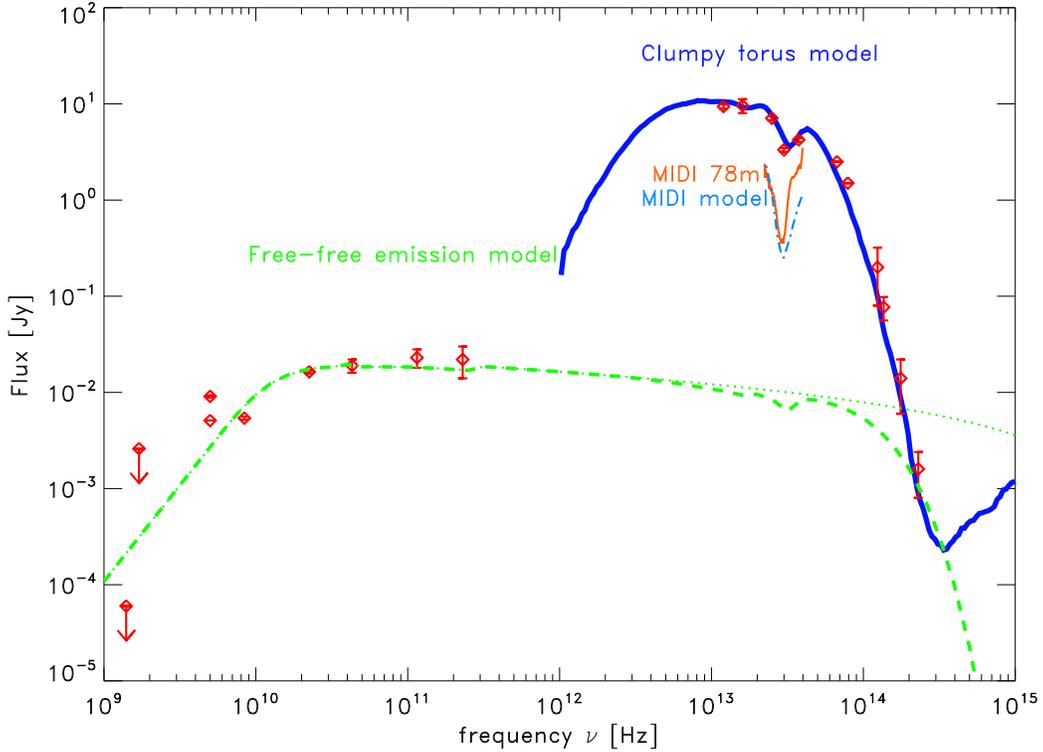}
\caption{Same as Fig.~\ref{Sync}, but with our free-free emission model in the radio. The free-free spectrum was attenuated by dust with $\tau_V=6$ to reproduce the $J$-band data.}\label{Free}
\end{figure*}

In a quantitative analysis, we inverted the low-frequency model approximation to synchrotron emission and self-absorption from quasi-monoenergetic electrons \citep{Bec97}. We suppose that the minimum energy assumption holds for our emission region to link $B$-field and electron density, $\nel$. The emission region is asummed to be a ellipsoid with apparent axes $r,s$ (volume $4/3 \pi r^2 s d^3$, and distance $d$ to NGC~1068; disk-like if $s<r$). In case the low-frequency turnover is due to SSA and the peak of the spectrum (end of the $\nu^{1/3}$-part) can be determined from observations, it is generally possible to fix the main model parameters, i.e. the $B$-field strength, the electron density $\nel$, and the Lorentz factor $\gamma$. Here, we don't know the peak of the spectrum and have upper limits to the self-absorption frequency since we assume that the steep low-frequency turnover is due to free-free absorption rather than SSA. Thus, only limits of the physical parameters can be determined.

For reference, we use the radio flux of the partly resolved core S1 at $\nu_1 =5\,$GHz from \citep{Gal04} of $F_{\nu_1} = 9.1\pm0.8\,$mJy.  The size of the elliptical source is $16.6$ and $11.2$\,mas for major and minor axis respectively. In Table \ref{tableSync} we use the half axis values corresponding to radii in the source model.

Here, we present three synchrotron models for S1. All are based on the $\nu^{1/3}$-power law spectrum as described above. In the first case (model A), we assume that the disk is resolved in the radial and vertical direction with a half major axis $r = 8.3\,$mas and a half minor axis $s = 5.6\,$mas. The corresponding source parameters are given in Table~\ref{tableSync}. In model B, we assume that the actual disk thickness is smaller than the minor axis, $s < 5.6\,$mas, so that the observed vertical extension might be due to disk inclination or warping. In this case we derive a lower limit for $s$ from the absence of SSA down to 3\,GHz. Thus, a lower limit for the mean Lorentz factor $\gamma$ of the electron distribution is obtained, as well as an upper limit for the corresponding $B$-field and also for the density of relativistic electrons in the radiating plasma (values see Table~\ref{tableSync}).

The altenative model C is motivated by the orientation of the major axis of S1 which is almost perpendicular to the large scale radio jet in NGC~1068
\citep{Mux96,Gal04}. S1 may, therefore, be associated with a perpendicular shock front close to the base of an outflow or the corona of an accretion disk.
The large size of S1 and its low brightness temperature of $T_{5\,{\rm GHz}} = 3.5\times10^6$\,K argues against a large scale shock. Instead, a corona above an accretion disk
may be heated in locallized regions through magnetic reconnection or small scale shock fronts. The corresponding emission regions are much smaller than the size of the underlying disk. In this model C, we assume $N \sim 100$ unresolved coronal emission regions which together form S1 and give it the appearence of a thick disk.
For simplicity we assume spherical emission regions and use a 3 times larger size than implied by the SSA limit to avoid mutual absorption by emission regions superimposed on the sky. We obtain lower limits to the sphere radius $s$ and to a mean Lorentz factor $\left<\gamma\right>$, and upper limits for $B$-field and relativistic electron density $\nel$. The results are summarized in Table~\ref{tableSync}. In this picture, the relatively large size of S1 as compared to the small emission regions might be due to electron scattering not directly connected to the small emission regions (e.g. surrounding medium) and provides only an upper size limit.

For NGC~1068, the possibility of synchrotron emission from quasi-monoenergetic electrons was already explored by \citet{Wit98} but with less radio data available. Similar models have been suggested for other galactic nuclei like Sgr A* \citep{Du94,Bec97} or M81 \citep{Reu96}. According to \citet{Wit04}, such an emission mechanism would require a source size of the order of 100\,AU and electron densities of $10^3\,{\rm cm^{-3}}$. While the size is consistent with our estimates, we require lower electron densities in the presented analysis.

\paragraph{Free-free emission.} The fluxes at the highest radio frequencies $>$100\,GHz are consistent with a flat spectrum. This opens the possibility that the radio spectrum is produced by thermal free-free emission, as already suggested by \citet{Gal97,Gal04}.

To investigate this possibility we setup a free-free emission model. The flux from thermal free-free emission at frequency $\nu$ for a given electron temperature $T$ is described by
\begin{equation}
\Snu = \Bnu(\nu,T) \left(1-\eexp^{-\tauff}\right)
\end{equation}
where $\Bnu(\nu,T)$ denotes the Planck function $\Bnu=2h\nu^3/c^2\cdot (\exp(h\nu/(kT)-1)^{-1}$. The optical depth for the free-free process can be described as
\begin{eqnarray}
\tauff & = & 1.13725 \ \left(1-\eexp^{-\frac{h\nu}{kT}}\right) \ \gaunt(\nu,T) \nonumber \\
   & & \times \left(\frac{T}{\rm K}\right)^{-1/2} \ \left(\frac{\nu}{\rm GHz}\right)^{-3} \ \left(\frac{\nel}{\rm cm^{-3}}\right)^{2} \ \left(\frac{\ell}{\rm pc}\right) \ .
\end{eqnarray}
Here, the emission measure $EM=\int \nel^2 \ {\rm d}\ell$ is approximated by a slab with constant electron density $\nel$ and a typical size $\ell$. The function $\gaunt$ is the Gaunt factor which accounts for the difference between semi-classical and quantum mechanical treatment of free-free emission \citep{Kra23,Gau30}. For low frequencies where $\gaunt>1$ we use the thermally averaged approximation by \citet{Ost61,Ost70},
\begin{equation}
\gaunt(\nu,T) = 1.5\ln\left(T/{\rm K}\right)-\ln\left(\nu/{\rm GHz}\right) - 5.307 \ . \label{gaunt_oster}
\end{equation}
For high frequencies, exact treatments show that $\gaunt\rightarrow1$ \citep[e.g.][]{Bec00} which we account for by a function that steps in when eq.~(\ref{gaunt_oster}) would become $\le 1$.

\begin{table}
\centering
\caption{Parameters for the free-free emission model as described in the text.}\label{Tab:ffpar}
\begin{tabular}{l c c c c c}
\\ \hline\hline
Model & $\ell/D$ & $\ell$ & $\nel$ & $T$ & $\tau_V$ \\
& [mas] & [pc] & [cm$^{-3}$] & [K] & \\ \hline
F & 11 & 0.75 & $8\times10^5$ & $1.3\times10^6$ & 6\\ \hline
\end{tabular}
\end{table}

In Fig.~\ref{Free} we show a free-free emission model fit to the radio SED of NGC~1068. The model parameters (noted as model F) are shown in Table~\ref{Tab:ffpar}. We attenutated the model with ISM dust of optical depth $\tau_V=6$ to be consistent with the NIR measurements, which concerns mainly the $J$-band data point in this case. The low-frequency cutoff occurs at the transition where the medium becomes optically thick to free-free emission. With out parameters, we obtain a cutoff frequency of $\nu(\tauff=1)=13.1\,{\rm GHz}$. For the size of the emission region, we used the resolved core diameter of 5 and 8.4\,GHz observations by \citet[][semi-major axis $\sim 11$\,mas]{Gal04}. These authors also provide parameters for a free-free model, based on fluxes at their observed frequencies, which are approximately in agreement with our model using a broader frequency range.

\section{The influence of the radio components to the IR}\label{Comp}

\begin{figure*}
\centering
\includegraphics[angle=0,width=14.5cm]{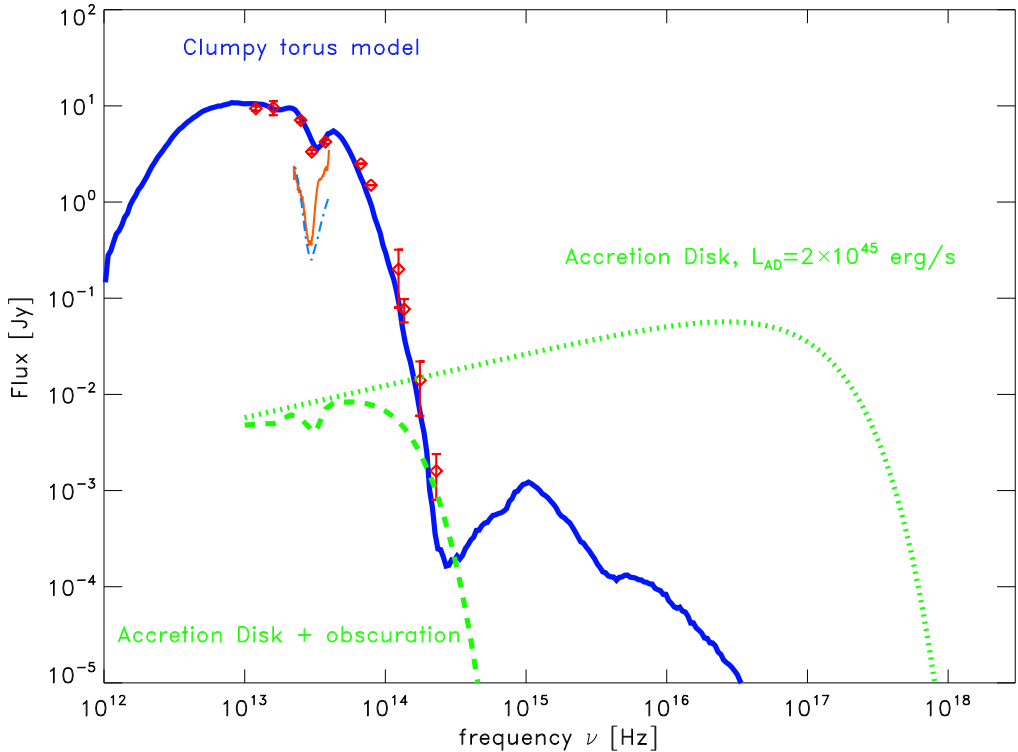}
\caption{Accretion Disk contribution to the IR spectrum. A standard accretion disk spectrum $F_\nu \propto \nu^{1/3}\exp(\nu/\nu_{\rm max})$ was adjusted for the assumed parameters $L=2\times10^{45}\,{\rm erg/s}$ and $M_{\rm BH}\approx10^7\,{\rm M_{\sun}}$ of NGC~1068 (green dotted line) and attenuated by dust obscuration (green dashed line) with $\tau_V=9.6$ to be consistent with the NIR measurements.}\label{ADspec}
\end{figure*}

Since there is a gap between the 100\,GHz and 10\,THz regime, it is difficult to judge how the IR and radio components are connected. In particular, a high frequency turnover for the radio component cannot be determined from the existing data. As a consequence, the radio source either breaks before the IR or reaches IR frequencies and, thus, may contribute to the unresolved IR fluxes according to our models as discussed in the previous section. The latter possibility leaves two open questions: (1) Is the unresolved flux in the long-baseline MIR interferometric data from VLTI/MIDI coming from the central AGN, i.e. the radio source? (2) Can the radio source (especially a possible synchrotron source) be responsible for a significant fraction of the unresolved $K$ band flux as observed by \citet{Wit04}? To address these questions, we extended the two investigated radio models into the IR regime, as discussed in the previous section (Figs~\ref{Sync} \& \ref{Free}, green dotted lines).

We assume that the radio source is surrounded by the dusty torus. As a result, it will suffer from extinction by the dust within the torus (Figs~\ref{Sync} \& \ref{Free}, green dashed lines). The extiction that were used to reproduce the $J$-band observation with the radio emission models are, however, much smaller than what is inferred from the lack of optical and infrared broad lines \citep[$A_V\ga50$\,mag][]{Lut00}. Thus, the actual IR contribution of the radio emission source is most likely less than what is derived from our approach to reproduce the $J$-band data. This enforces the analysis in terms of obtaining strong upper limits for the contribution to the unresolved $K$-band and mid-IR flux.

As can be seen from Figs~\ref{Sync} \& \ref{Free}, the maximum possible extrapolated mid-IR flux of the synchrotron and free-free power law is still a factor of 5-10 below the unresolved 78\,m baseline MIDI flux. As a conservative limit, we estimate the maximum flux contribution of the radio source to the MIR $F_{\rm Radio}(MIR)/F(MIR) \ll 0.2$ if it is sychrotron emission. This means that the unresolved MIDI flux is most likely originating exclusively from thermal dust emission in the torus rather than from the radio component. As shown in the previous section \citep[see also][]{Hon06,Hon07a}, the unresolved emission can be fully interpreted in terms of a clumpy torus model.

In the near-IR, the constraints are even tighter since (1) the assumption that 100\% of the $J$-band flux is very conservative and (2) the difference in ``true'' (= broad-line region) and modelled extinction affects near-IR frequencies stronger than mid-IR frequencies. As can be seen in Figs~\ref{Sync} \& \ref{Free}, the non-torus contribution to the $K$-band emission is certainly much less than 20\%, no matter if the radio emission is due to synchrotron or free-free emission. This very strong upper limit is important for the interpretation of the unresolved flux measured by \citet{Wit04} using VLTI/VINCI: Even though the size of this unresoved emission is $\le0.3$\,pc, it must originate from the torus. \citet{Kis07} show that this size is consistent with the sublimation radius as expected from reverberation mapping of type 1 objects and assuming a central AGN luminosity of $10^{45}$\,erg/s. As an alternative, substructure in a clumpy torus can also account for the unresolved flux \citep{Hon07a}.

\section{Quantifying the contribution of the accretion disk}\label{AD}


An alternative interpretation for the unresolved VLTI/VINCI flux is contribution from the putative accretion disk in the center of NGC~1068, as discussed by \citet{Wit04}. In that case, the torus requires to be clumpy so that the accretion disk can be seen through ``holes'' of low or no extinction within the torus.

In a clumpy medium with high optical depth of each dust cloud, the probability, $P_{\rm AD}$, that an accretion disk photon crosses all clouds without being absorbed can be estimated by Poisson statistics, $P_{\rm AD}=\exp(-N)$ \citep[$N$ = number of clouds along the line-of-sight;][]{Nat84}. From the edge-on torus model as presented in Sect.~\ref{Models}, we obtained $N = 10$. This results in a probability to see the accretion disk through the torus clouds of $P_{\rm AD}\sim10^{-5}$, which makes the scenario very unlikely. Nevertheless, we aim for constraining the possible accretion disk contribution to the $K$-band flux. For that, we used a standard accretion disk spectrum of the shape $F_\nu \propto \nu^{1/3}\exp(\nu/\nu_{\rm max})$ \citep[e.g.,][]{Sha73}. The turnover frequency, $\nu_{\rm max}$, is characterised by the accretion disk temperature at the innermost stable orbit ($\sim 3R_{\rm Schwarzschild}$), which can be inferred from the mass of the central black hole \citep[$M_{\rm BH}\approx10^7\,{\rm M_{\sun}}$,][]{Gre96} and the modelled bolometric luminosity ($L=2\times10^{45}\,{\rm erg/s}$, see Sect.~\ref{IRModel}). For NGC~1068, we obtain $\nu_{\rm max}\sim 8\times10^{16}\,{\rm Hz}$ which corresponds to a temperature of approximately $1.5\times10^6\,{\rm K}$. We extended the accretion disk spectrum down to IR wavelengths, so that the integrated accretion disk spectrum equals the bolometric luminosity as inferred from torus modelling. This allows to get an idea of the thermal emission from the accretion disk, although it is completely obscured in the UV and optical.

In Fig.~\ref{ADspec}, we show a comparison of the observed fluxes with the clumpy torus model (Sect.~\ref{IRModel}), and the accretion disk spectrum which is extrapolated from the UV to the MIR (dotted line) and attenuated by dust extinction with $\tau_V=9.6$ (dashed line) to reproduce the $J$-band flux. The obscured accretion disk spectrum is obviously not able contribute significantly to the NIR or MIR emission. Again, it should be noted that the actual extinction towards the broad line region is much higher than used to obtain the upper limit as described in the previous section. Thus, we safely conclude that that the unresolved $K$-band flux as observed with VLTI/VINCI (40\% of the total nuclear flux) does not originate from the accretion disk.


\section{Summary and Conclusions}\label{Conc}

We analysed the SED of NGC\,1068 from $10^9$\,Hz to $10^{14}$\,Hz. The SED was modelled by two components: (1) an IR component which is represented by a clumpy torus model, and (2) a radio component represented either by synchrotron emission with $F_\nu \propto \nu^{1/3}$, or optically thin free-free emission. For the infrared, we used our clumpy torus model to simultaneously reproduce the high-resolution NIR-to-MIR SED and the VLTI/MIDI correlated fluxes at 78\,m baseline. The model is characterized by a radial cloud density distribution $\propto r^{-3/2}$ and a half opening angle of 40$\degr$. The central AGN is obscured by 10 dust clouds on average along a radial line of sight in the equatorial plane. For the radio source, we showed that the power-law index of $0.29\pm0.07$ is consistent with a spectral slope proportional to $\nu^{1/3}$ which is expected from synchrotron emission by quasi-monoenergetic electrons. Further more, we showed that the radio emission can also originate from thermal free-free emission. Both radio models require a low-frequency turnover between 1 and 10\,GHz where the spectral slope changes. The steepness in the range between 1-6\,GHz suggests that the turnover is caused by a foreground screen of thermal plasma which becomes optically thick below $\sim$3-6\,GHz ($F_\nu \propto \exp\left(-(\nu/3\,{\rm GHz})^{-2.1}\right)$). This can be either connected to the gas in the galaxy or in the torus.

For both scenarios -- synchrotron and free-free emission -- we presented models to extract physical parameters from the radio SED, depending on the emission mechanism. The observed source sizes at 5 and 8.4 GHz are in agreement with synchrotron emission coming either from a disk (models A \& B) or from small emission regions connected to the disk (e.g. magnetical reconnection zones in the disk corona) and possibly additional electron scattering in the surrounding medium (model C). The necessary Lorentz factors are $\gamma>10^4$ with $B$-field strengths of the order of $0.1-1$\,G. In case the radio SED is due to thermal free-free emission (model F), the observed fluxes and source sizes require an electron density $\nel=8\times10^5\,{\rm cm^{-3}}$ and a temperature $T=1.3\times10^6\,{\rm K}$.

The main goal of this study was to spectrally disentangle the near- and mid-infrared torus dust emission from possible contamination by the radio source. For that, we extrapolated the free-free and synchrotron models to the infrared. It was possible to show that the radio source is contributing certainly $<$20\% to the observed unresolved flux at 78\,m baseline. A similar analysis shows that the unresolved $K$-band flux as observed with VLTI/VINCI is originating from the torus and almost uncontaminated by the radio source. We were also able to argue against the possibility that in the $K$-band the accretion disk might be seen through holes in the clumpy torus. This means that all observed near- and mid-IR emission -- resolved and unresolved -- is thermal dust emission from the torus. This interpretation is consistent with the clumpy torus model that was presented. It is, however, not clear if the unresolved $K$-band flux represents emission from the inner boundary of the torus as indicated by \citet{Kis07}, or due to the clumpy substructure of the torus \citep{Hon06,Hon07b}.

\begin{acknowledgements}
The authors would like to thank M.~Kishimoto for fruitful discussions and Ian Robson for helpful comments and suggestions that improved the paper.
\end{acknowledgements}

\bibliographystyle{aa}

\end{document}